\documentstyle[prl,multicol,aps]{revtex}
\begin{document}
\title{Spin-Charge Separation in Quantum Hall Liquids}
\author{H. C. Lee}
\address{Asia Pacific Center for Theoretical Physics,
Seoul, Korea}
\author{S.-R. Eric Yang}
\address{Department of Physics, Korea University, Seoul 136-701, Korea}
\maketitle
\draft
\begin{abstract}
We have investigated chiral edges of a quantum Hall(QH) liquid
at filling factor $\nu=2$.
We find 
that spin and charge separate in the presence of the long-range 
Coulomb interaction, and the tunneling density of states(DOS)
is given by $D(\omega)\sim [-1/\ln\omega]^{1/2}$.  
The measurement of the temperature and voltage dependences of the tunneling
current should reveal the presence spin-charge separation.
\end{abstract}
\thispagestyle{empty}
\begin{multicols}{2}

Spin charge separation, the transmutation of
statistics, and Luttinger liquid behavior are the fundamental properties of 
interacting 1D electrons \cite{E,S,H}.
Edge electrons of a QH liquid form an interacting 1D chiral
liquid \cite{W1,W2,W3,KF,MYKGF,FLS}.  For short-range interactions 
the electrons form a Luttinger
liquid at fractional filling factors and
a Fermi liquid at integer filling factors.
The experimental evidence of the power law behavior of 
Luttinger liquid 
came from tunneling between  two
$\nu=1/3$ edges \cite {MUW} and 
between a bulk doped-GaAs normal metal and the abrupt edge of a 
QH fluid \cite{CPW}.

The separation of spin and charge is an 
enormous simplification, and it also has important consequences for the 
low-energy physics.  
The wave functions and space-time correlation functions 
factorize into products and the spin and charge collective modes also 
have different velocities.  
However, spin charge separation has received little attention in connection
with QH edge electrons.
The edges can have a composite structure in the fractional
case\cite{W1,M} and 
spin charge separation
is not expected.  
We propose in this paper that  
the tunneling DOS should reveal spin charge separation at 
the integer case $\nu=2$, provided that the long range of the 
Coulomb interaction is present.  
The tunneling DOS is of the form $D(\omega)\sim [-1/\ln\omega]^{1/2}$  
and differs from 
$D(\omega)\sim [-1/\ln\omega]$ of the spin-polarized case $\nu=1$ \cite{FB}.  
The appearance of the power $1/2$ is a
consequence of spin charge separation and depends crucially on
the long range of the Coulomb interaction \cite{OF}. 
We have calculated 
the temperature and voltage dependences of the tunneling
current between a bulk doped-GaAs and the abrupt edge of a QH fluid.

In our model electrons reside on a rectangle of width $W$ and length $L$.
The single particle states may be denoted by a wavevector 
$k$ in the Landau gauge.  A periodic
boundary condition is used so that $k=\frac{2\pi}{L}n$, where $n$ is an
integer.
We assume that a finite edge separation
is present so that the Fermi velocities of spin up and down electrons are
different \cite{DGH}.  
Since the distance between the edges
of different chirality is large the coupling between them may be neglected.
We can center the coordinate 
system on the edge of spin down electrons
so that the Fermi wavevector of spin down electrons $k_{F,\downarrow}$ is zero.
In this model we assume that wavevectors can take values 
from $-\infty$ \cite{deW}.
We calculate the DOS of two edges of the same chirality
using a bosonization approach.  

The effective Hamiltonian\cite{W2} of edge electrons 
interacting via the long-range
Coulomb interaction 
has the form
\begin{eqnarray}
& &H=\sum_{s,p>0}\frac{2\pi v_s}{e^2}\rho_{s,p}\rho_{s,-p}
+\sum_{s,p>0}V_a(p)\rho_{s,p}\rho_{s,-p}\nonumber\\
& &+\sum_{s,p>0}V_b (p)\rho_{s,p}\rho_{-s,-p},
\end{eqnarray}
where the density operators satisfy the commutation relation 
$[\rho_{s,p},\rho_{s,-p^{\prime}}]=\frac{p e^2}{2\pi}\,\delta_{p,p^{\prime}}$.
The intra and inter edge coupling constants are given by
$V_a(p)=\frac{2}{\epsilon}\ln\frac{2}{|p|a}$ and
$V_b (p)=\frac{2}{\epsilon}\ln\frac{2}{|p|b}$. 
The Fermi velocity of the edge with spin $s$ is $v_s$.
The actual values of the physical parameters appearing in the Hamiltonian
must be determined
from a microscopic consideration \cite{ZM}.  
We form a new basis by taking a canonical transformation
$\rho_p=\frac{1}{\sqrt{2}}
\Big[ \rho_{\uparrow,p}+\rho_{\downarrow,p} \Big]$ and 
$\sigma_p=\frac{1}{\sqrt{2}}\Big[\rho_{\uparrow,p}-
\rho_{\downarrow,p} \Big]$.  The Hamiltonian reads
\begin{eqnarray}
H&=& \frac{2\pi}{e^2}\sum_{p>0} v_{c,p}:\rho_p \rho_{-p}:
+\frac{2\pi}{e^2}\sum_{p>0}v_{s,p} : \sigma_p\sigma_{-p}:\nonumber\\
& &+ \frac{2\pi}{e^2}\sum_{p>0}  g\sigma_p \rho_{-p}.
\end{eqnarray}
The renormalized charge and spin velocities are
$v_{c,p}=v+\frac{2e^2}{\pi \epsilon}\ln\frac{2}{|p|c}$ and 
$v_{s,p}=v+\frac{e^2}{\pi \epsilon}\ln\frac{b}{a}$, where $c^2=ab$ 
and $v=(v_{F,\uparrow}+v_{F,\downarrow})/2$.
The coupling between spin and charge modes is given by 
$g=v_{F,\uparrow}-v_{F,\downarrow}$.
This Hamiltonian can be diagonalized by the canonical transformation
$\rho_p=\cos \theta_p \tilde{\rho}_p-\sin \theta_p \tilde{\sigma}_p$ 
and $\sigma_p=\sin \theta_p\tilde{\rho}_p+\cos \theta_p\tilde{\sigma}_p$,
where $\tan2\theta_p=2g/\pi(v_{c,p}-v_{s,p})$.
The diagonalized Hamiltonian is
\begin{eqnarray}
H= \frac{2\pi}{e^2}\sum_{p>0}v^{'}_{c,p}:\tilde{\rho_p}\tilde{\rho}_{-p}:
+\frac{2\pi}{e^2}\sum_{p>0}v^{'}_{s,p} : \tilde{\sigma}_p\tilde{\sigma}_{-p}:,
\end{eqnarray}
where $v^{'}_{c,p}$ and $v^{'}_{s,p}$ are the renormalized velocities.
The bosonization formula for spin s is given by
\begin{eqnarray}
\psi_s(x)=\frac{e^{ik_{F,s} x}}
{\sqrt{2\pi \alpha}} \exp(-\frac{2\pi}{L}\sum_{p\neq 0}
\frac{e^{-\alpha |p|/2-ipx}}{p}\rho_{p,s}).
\end{eqnarray}
We find that the electron Green's function of spin $s$ is given by
\begin{eqnarray}
& &G_s(x,t)\sim \frac{e^{ik_{F,s} x}}{2\pi \alpha}
 \exp\Bigg[-\frac{1}{2}\,
\int_0^{\infty}\,\frac{d p}{p}\,e^{-\alpha p}\times \nonumber\\
& &\Big(2-e^{ip(x- v^{'}_{c,p}t)}
-e^{ip(x-v^{'}_{s,p} t)}) \Big)\nonumber\\ 
& &-\frac{s}{2}\,\int_0^{\infty}\,\frac{d p}{p}\,e^{-\alpha p}\sin(2\theta_p)
\Big(e^{ip(x-v^{'}_{s,p} t)}-e^{ip(x-v^{'}_{c,p}t)} \big) \Bigg].
\end{eqnarray}
For $\omega<<\omega_0$ we find
\begin{eqnarray}
D_s(\omega) \sim \frac{1}{\sqrt{\ln|\frac{1}{\omega}|}}\,\exp\Big[\frac{s g}{\pi
v_0}/\ln \left|\frac{1}{\omega}\right| \Big],
\end{eqnarray}
where $\omega_0$ is of order
$e^2/\epsilon W$, $v_0=2e^2/(\epsilon\pi)$, and $\omega$ 
is measured from the Fermi energy.
This result should be contrasted to the result
$D(\omega)\sim [-1/\ln\omega]$ at $\nu=1$.  
The square root in the denominator reflects spin-charge separation.

The groundstates at $\nu=1,2$ are both exactly given by
HF theory, and it is illustrative to consider the DOS from the perspsective 
of HF theory.
It is sufficient to consider the limit where
the confinement potential is very steep so that the groundstate is given
by a single Slater determinant with $N_\uparrow$ and $N_\downarrow$
spin-up and down electrons
$|N_\uparrow, N_\downarrow,0>$ \cite{YMJ,MYJ}.
This state
has, unlike a normal Fermi liquid, zero
mass renormalization in spite of the strong electron-electron 
interactions.
According to the Lehmann representation of the Green's function  
\begin{eqnarray}
G_s(k,\omega)&=&\sum_n (\frac{|<N_\uparrow+1,N_\downarrow,n|
c_{k,\uparrow}^+|N_\uparrow,N_\downarrow,0>|^2}{\omega-(E_n(N_\uparrow+1,N_\downarrow)
-E_0(N_\uparrow,N_\downarrow))}\nonumber\\
&+&\frac{|<N_\uparrow-1,N_\downarrow,n|
c_{k,\uparrow}|N_\uparrow,N_\downarrow,0>|^2
}{\omega-(E_0(N_\uparrow,N_\downarrow)-
E_n(N_\uparrow-1,N_\downarrow))}).
\end{eqnarray}
Here $|N_\uparrow,N_\downarrow,n>$ 
and $E_n(N_\uparrow,N_\downarrow)$ are the n'th lowest energy state and
its energy. 

When $\nu=1$ ($N_\downarrow=0$) 
it is possible to compute
several values of the self energy exactly within this HF theory.
Consider the first term in the Lehmann representation.
Since the groundstate $|N_\uparrow+1,0,0>=c_{k_F+\Delta k,\uparrow}^+
|N_\uparrow,0,0>$ (Fig.1 (a))
and the first excited state
$|N_\uparrow+1,0,1>=c_{k_F+2\Delta k,\uparrow}^+
|N_\uparrow,0,0>$ (Fig.1 (b)) are single Slater determinant states
the self energies $\Sigma_{k_{F,\uparrow+\Delta k}}$ 
and $\Sigma_{k_{F,\uparrow}+2\Delta k}$ are exactly given by the HF theory.  
Now consider the second term in the Lehmann representation.
The self energy 
$\Sigma_{k_{F,\uparrow}}$ is 
also known exactly since $|N_\uparrow-1,0,0>=c_{k_F,\uparrow}
|N_\uparrow,0,0>$ (Fig.1 (c)) is a single Slater determinant state.
>From these values of the self energies the qualitative 
behavior of the DOS may be extracted.
It is also possible to calculate the asymptotic values of the
self energy directly. 
In the limit $k\rightarrow k_F$ 
the first derivative of the exchange self energy  
has a singularity at $k= k_F$.
>From the expression of the exchange self energy
\begin{eqnarray}
\Sigma_{k,s}^X=-\frac{e^2}{2\pi \epsilon}\int_{-\infty}^{k_{F,s}}dk'
e^{-(\frac{k-k'}{2})^2}K_0 ((\frac{k-k'}{2})^2),
\end{eqnarray}
we find
$\Sigma_{k,s}^X \sim (k-k_F) ln (k-k_F)$ ($K_0$ is the modified Bessel 
function).
>From this result we can rederive the bosonization result
$D(\omega)\sim -1/\ln \omega$ in the limit $\omega\rightarrow 0$\cite{FB}.
Although the groundstate at $\nu=2$ is exactly given by a single Slater 
determinant
the low energy behavior of the DOS 
cannot be derived from the HF theory.
The main difference from the $\nu=1$ case is that the first excited
state $|N_\uparrow+1,N_\downarrow,1>$ 
is a linear combination
of two Slater determinants 
$c^+_{k_{F,\downarrow}+\Delta k}c_{k_{F,\downarrow}}
c^+_{k_{F,\uparrow+\Delta k}}
|N_\uparrow,N_\downarrow,0>$ (Fig.2 (c)) 
and $c^+_{k_{F,\uparrow}+2\Delta k} |N_\uparrow,N_\downarrow,0>$ (Fig.2 (d)).
The HF theory thus cannot give the correct self energy at
$\Sigma_{k_{F,\uparrow}+2\Delta k}$.  

It is illustrative to demonstrate microscopically 
that charge and spin edge modes have different
velocities.  Since the edge of a rectangular sample is equivalent to the
edge of a large quantum dot
we will consider a dot with $N_\uparrow=N_\downarrow=N$. 
In the symmetric gauge the single particle states may be labeled by
the angular momentum component along the magnetic field.
At angular momentum $k=0$
spin and charge collective modes have the same energy.  
At the next highest angular momentum
$k=\Delta k=1$
the relevant many body Hilbert space may be expanded
by two single Slater determinant states 
$|\phi_{\uparrow}>=c^+_{k_{F}+\Delta k,\uparrow}
c_{k_{F},\uparrow}|N,N,0>$
and $|\phi_{\downarrow}>=c^+_{k_{F}+\Delta k,\downarrow}
c_{k_{F},\downarrow}|N,N,0>$.
The diagonal matrix elements of the 2x2 Hamiltonian are
\begin{eqnarray}
H_d=\epsilon_{k_{F}+\Delta k}
-\epsilon_{k_{F}}-D_{k_{F},k_{F}+\Delta k}+X_{k_{F},k_{F}+\Delta k},
\end{eqnarray}
where $\epsilon_{k}$ is the quasiparticle energy renormalized
by the self energy and the third and fourth terms represent
the excitonic and depolarization vertex corrections.
The off-diagonal elements are 
\begin{eqnarray}
H_o=D_{k_{F},k_{F}+\Delta k}.
\end{eqnarray}
The quantities $D_{m,m'}$ and $X_{m,m'}$ are the direct and exchange Coulomb
matrix elements.
The charge eigenmode $\phi_{\uparrow}+\phi_{\downarrow}$ has energy 
\begin{eqnarray}
E_C=\epsilon_{k_{F}+\Delta k}
-\epsilon_{k_{F}}+X_{k_{F},k_{F}+\Delta k},
\end{eqnarray}
and
the spin eigenmode $\phi_{\uparrow}-\phi_{\downarrow}$ has energy
\begin{eqnarray}
E_S=\epsilon_{k_{F}+\Delta k}
-\epsilon_{k_{F}}-2D_{k_{F},k_{F}+\Delta k}+X_{k_{F},k_{F}+\Delta k}.
\end{eqnarray}
Since the spin and charge modes have the same energy at $k=0$ but different
energies at $k=\Delta k$ they must have different velocities.

We will now investigate the DOS when the filling factor is $\nu=4/3$.
The filling factor of spin-up and -down electrons are $\nu=1$ and $\nu=1/3$.
Including the coupling between the edges within Wen's effective Hamiltonian
we find 
\begin{eqnarray}
D_{\uparrow}(\omega)\sim1/\ln|1/\omega|,
\end{eqnarray}
and
\begin{eqnarray}
D_{\downarrow}(\omega)\sim |\omega|^{\frac{1}{\nu}-1}\frac{1}
{|\ln1/\omega|^{1/\nu}}.
\end{eqnarray}
These results for DOS are identical to those of isolated
edges in the absence of 
edge coupling, and spin-charge is thus absent at $\nu=4/3$. 
It should be stressed that the logarithmic corrections
to the DOS are due to the long range 
of the Coulomb interaction.  

Experimentally relevant quantities are the voltage and temperature
dependences of
the tunneling current.
We calculate these quantities for the structure used in a recent
experiment \cite{CPW}, where a AlGaAs tunnel barrier is inserted between
a 2D electron gas and a 3D doped GaAs.
For $eV$ and $T$ less than $\omega_0$ it is possible to obtain
simple analytical results, and the results are given below.
For $\nu=2$ and $T=0$ 
we find, for both spin channels, that the voltage dependence is
given by
\begin{eqnarray}
I \propto \frac{eV}{\Big(\ln \frac{1}{eV}\Big)^{1/2}}.
\end{eqnarray}  
The temperature dependence is given by the following approximate interpolating
expression between the low and high temperature limits
\begin{eqnarray}
I \propto   eV 
\frac{1}{\Big( \ln \frac{1}{\max ( T, eV/2)} \Big)^{1/2}}.
\end{eqnarray}
The dominant tunneling current arises from tunneling into spin up
states which are closest to the 3D reservoir.
The edge separation between
spin up and down electrons is typically of the order of the magnetic length.
We have also calculated the voltage and temperature dependences 
at $\nu=1$:
\begin{eqnarray}
I_{\uparrow} \propto \frac{eV}{\ln \frac{1}{eV}},
\end{eqnarray}
and 
\begin{eqnarray}
I_{\uparrow} \propto  eV
\frac{1}{\ln \frac{1}{\max ( T, eV/2)} }.
\end{eqnarray}
For comparison, the voltage dependence is calculated for $\nu=1/3$  down spin edge.
\begin{eqnarray}
I_{\downarrow} \propto \nu \frac{(eV)^{1/\nu}}
{\Big|\ln \frac{1}{eV} \Big
|^{1/\nu}},
\end{eqnarray}
The temperature dependence is given by
the following approximate interpolating expression 
between the low and high temperature limits
\begin{equation}
I_{\downarrow} \propto T^{1/\nu}
\,\Bigg[ \frac{eV}{k_B T}\,\frac{1}{\big|\ln \frac{1}{T}\big|^{1/\nu}}+
\left( \frac{eV}{k_B T} \right)^{1/\nu}\,
\frac{1}{\big|\ln \frac{1}{eV}\big|^{1/\nu}}\Bigg].
\end{equation}
These results for $\nu=1/3$ are not identical to the previous
results for short range interactions: the
Coulomb interaction gives arise to the logarithmic corrections.
To observe the
logarithmic corrections the experimental data 
should be fitted to this expression over a wider range of $T$ and $V$.

A QH chiral liquid at $\nu=2$ is a rather unique 1D liquid.
Although it is not a Luttinger liquid spin charge separation is present
provided the electron-electron interactions are given
by the long range Coulomb interaction.
Our work shows that this effect can be observed in  
the tunneling current between   
a bulk doped-GaAs and the abrupt edge of a QH fluid \cite{CPW}.

This work has been supported by the Non Directed Researh Fund (Korea Research
Foundation), the KOSEF
under grant 961-0207-040-2, and the Ministry of Education under grant
BSRI-96-2444.

\end{multicols}

\begin{figure}
\caption{Occupation numbers of the groundstate $|N_{\uparrow}+1,0,0>= 
c^+_{k_{F,\uparrow}+\Delta k}|N_{\uparrow},0,0>$, and the first excited 
state $|N_{\uparrow}+1,0,1>=
c^+_{k_{F,\uparrow}+2\Delta k}|N_{\uparrow},0,0>$ are shown in (a) and (b).
The occupation number of the groundstate of $N_{\uparrow}-1$ electrons,
$|N_{\uparrow}-1,0,0>=
c_{k_{F,\uparrow}}|N_{\uparrow},0,0>$, is shown in
(c).  At $\nu=2$ the first excited state $|N_{\uparrow}+1,N_{\downarrow},1>$
is a linear combination of two Slater determinant states
$c^+_{k_{F,\downarrow}+\Delta k}c_{k_{F,\downarrow}}c^+_{k_{F,\uparrow}}
|N_{\uparrow},N_{\downarrow},0>$ and 
$c^+_{k_{F,\uparrow}+2\Delta k}|N_{\uparrow},N_{\downarrow},0>$.  
The occupation numbers of these two states are
shown in
(d), and (e).}
\end{figure}
\end{document}